\documentclass[aps,preprint]{revtex4}%
\usepackage{amsfonts}
\usepackage{amsmath}
\usepackage{amssymb}
\usepackage{graphicx}%
\begin{document}
\title[BLACKBODY RADIATION AND ZERO-POINT RADIATION]{CONNECTING BLACKBODY RADIATION AND ZERO-POINT RADIATION WITHIN CLASSICAL
PHYSICS: A NEW MINIMUM PRINCIPLE AND A STATUS REVIEW}
\author{Timothy H. Boyer}
\affiliation{Department of Physics, City College of the City University of New York, New
York, NY 10031}

\begin{abstract}
A new thermodynamic analysis is presented for the intimate connections between
blackbody radiation and zero-point radiation within classical physics. First,
we discuss the thermodynamics of a classical harmonic oscillator. Using the
behavior under an adiabatic change of frequency, we show that the
thermodynamic functions can all be derived from a single function of
$\omega/T$, analogous to Wien's displacement theorem. The high- and
low-frequency limits allow asymptotic energy forms involving $T$ alone or
$\omega$ alone, corresponding to energy equipartition and zero-point energy.
It is then suggested that the actual thermodynamic behavior for a harmonic
oscillator is given by the function satisfying the Wien displacement result
which provides the smoothest possible interpolation between scale-decoupled
energy equipartition at low frequency and scale-invariant zero-point energy at
high frequency. This suggestion leads to the Planck spectrum with zero-point
energy. Second, we turn to radiation in a box in one space dimension. We
consider a box with conducting walls and a conducting partition so that the
discrete normal mode structure of the box becomes important. The contrasting
Casimir energies are explored for the Rayleigh-Jeans and zero-point spectra.
The Rayleigh-Jeans spectrum involves no change of energy with partition
position, and the zero-point spectrum involves no change of entropy. \ It is
suggested that the Planck spectrum with zero-point radiation satisfies a
natural minimum principle which corresponds to greatest independence of the
system energy from the position of the partition for a fixed temperature.
Numerical calculation shows that the familiar spectra other than the Planck
spectrum violate the minimum principle. It is also remarked that in the
absence of zero-point radiation, there is no finite energy minimum solution
and the familiar "ultra-violet divergence" reappears. Third, we review the
previous derivations of the Planck radiation spectrum in classical physics,
all of which involve zero-point radiation. We again suggest, based upon
scaling symmetry, that only purely electromagnetic scattering systems will
lead to thermal equilibrium involving finite thermal radiation energy above
the zero-point radiation spectrum.

\end{abstract}
\maketitle

\bigskip\draft

\noindent\textbf{INTRODUCTION}

Discussions of blackbody radiation rarely make reference to zero-point
radiation. Within quantum physics textbooks, blackbody radiation is regarded
as comprehensible only in terms of quantum statistical mechanics, and quantum
statistical mechanics has no role for zero-point radiation. However, within
classical physics, the two spectra are intimately related; zero-point
radiation provides the basis for understanding the blackbody spectrum. This
has been emphasized in the past, and in this article we describe a new aspect
of this relation. It is pointed out that the thermal radiation spectrum is the
smoothest possible transition from the scale-decoupled Rayleigh-Jeans spectrum
at low frequency to the scale-invariant zero-point radiation spectrum at high
frequency. In this article we will provide a quantitative meaning to this
"smoothest" criterion while providing a review of the connections between
blackbody radiation and zero-point radiation within purely classical physics.

\noindent\textbf{OUTLINE OF THE PRESENTATION}

Our presentation is divided into three basic parts. The first part reviews the
thermodynamics of the classical harmonic oscillator and obtains the energy
functional dependence $\mathcal{U}(T,\omega)=\omega f(\omega/T)$ which is
associated with the Wien displacement law. We note that zero-point energy and
energy equipartition represent the opposite extremes of this theorem, one
energy spectrum involving frequency alone and the other temperature alone.
Next we suggest that the actual thermodynamic behavior is given by the
smoothest possible interpolation between the equipartition asymptotic form at
low frequency and the zero-point energy asymptotic form at high frequency.
Using this demand for a smooth interpolation, we give a suggestive derivation
of the Planck spectrum with zero-point radiation.

In the second part of this paper, we turn from a single oscillator over to
thermal radiation. Because of the greater ease of calculations, we carry out
the analysis for waves in one spatial dimension rather than three dimensions.
We first obtain the Stefan-Boltzmann law for the energy density of thermal
radiation in a very large one-dimensional box. Then we discuss the
thermodynamics of a one-dimensional system involving Casimir forces on a
partition which divides the box into two compartments, and we point out that
any spectrum of random classical radiation will lead to Casimir forces. Then
we obtain the Casimir energies associated with the zero-point spectrum and the
Rayleigh-Jeans spectrum. The zero-point radiation spectrum is special because
it is the unique spectrum which involves no dependence of the entropy upon the
position of the Casimir partition; the Rayleigh-Jeans spectrum is special
because it is the unique spectrum which involves no dependence of the energy
upon the position of the Casimir partition. From numerical calculations, it is
easily seen that thermodynamic principles applied to the Casimir thermodynamic
system lead to restrictions on the allowed thermal radiation spectra. We then
explore a natural minimum principle for the Casimir energies of various
spectra. We find the traditional "ultra-violet catastrophe" when classical
zero-point energy is not included in the minimum principle. When zero-point
energy is included, then numerical calculation suggests that Planck's spectrum
satisfies the minimum principle whereas other spectra do not. We conjecture
that analytic solution of the minimum principle would confirm the Planck spectrum.

In the third part of this paper, we review the information in the literature
regarding blackbody radiation within classical physics, noting the variety of
classical derivations of the Planck spectrum, all of which depend upon the
presence of zero-point radiation. Also, we note the contradictory evidence
regarding scattering by general classical mechanical systems. In this
connection we again suggest that only purely electromagnetic systems have the
possibility of giving the Planck spectrum with zero-point radiation as the
equilibrium radiation spectrum. Finally we make some comments on classical
theory within contemporary physics.

\noindent\textbf{PART I: THERMODYNAMICS OF A HARMONIC OSCILLATOR}

\noindent\textbf{DERIVATION OF FUNCTIONAL DEPENDENCE CORRESPONDING TO WIEN'S
DISPLACEMENT THEOREM}

As a beginning step in discussing thermal radiation, we consider the
thermodynamics of a classical harmonic oscillator system. We first need to
recognize the generalized force associated with a change in the natural
oscillator frequency. We consider a simple harmonic oscillator system with
natural frequency $\omega$ described by the Lagrangian $L(q,\dot{q})$ and
Hamiltonian $H(q,p)$
\begin{equation}
L(q,\dot{q})=(1/2)m\dot{q}^{2}-(1/2)m\omega^{2}q^{2},\;\;~~~H(q,p)=p^{2}%
/(2m)+(1/2)m\omega^{2}q^{2}%
\end{equation}
This system can also be described in terms of action-angle variables
$\tilde{J}$, $\tilde{w}$ where
\begin{equation}
q=\left(  \frac{2\tilde{J}}{m\omega}\right)  ^{1/2}\sin(\tilde{w}%
),\;\;\;\;p=(2m\tilde{J}\omega)^{1/2}\cos(\tilde{w})
\end{equation}
and the Hamiltonian is $H(\tilde{J})=\tilde{J}\omega$. [1] \ Now we are
interested in the change of energy of the oscillator under a very slow
alteration of the oscillator's natural frequency $\omega$. During such a slow
alteration, the action variable $\tilde{J}$ is constant.[2] Thus the change of
energy of the oscillator is
\begin{equation}
dH=\tilde{J}d\omega=(H/\omega)d\omega
\end{equation}
and the work done by the oscillator on the agent altering $\omega$ is just the
negative of this
\begin{equation}
dW=-(H/\omega)d\omega.
\end{equation}
Thus for an oscillator of energy $\mathcal{U}$ and (angular) frequency
$\omega$, we regard $\mathcal{X}=-\mathcal{U}/\omega$ as the generalized force
associated with a change in $\omega$.

When the oscillator is at thermal equilibrium in a bath at temperature $T$,
the oscillator will have an (average) energy $\mathcal{U}(T,\omega)$ and an
entropy $\mathcal{S}(T,\omega)$ depending upon the temperature $T$ and natural
oscillation frequency $\omega$. Using the work expression in(4), the laws of
thermodynamics for the oscillator give
\begin{equation}
dQ=Td\mathcal{S}(T,\omega)=d\mathcal{U}(T,\omega)-(\mathcal{U}/\omega)d\omega
\end{equation}
This can be rewritten as
\begin{equation}
T\left[  \left(  \frac{\partial\mathcal{S}}{\partial T}\right)  _{\omega
}dT+\left(  \frac{\partial\mathcal{S}}{\partial\omega}\right)  _{T}%
d\omega\right]  =\left(  \frac{\partial\mathcal{U}}{\partial T}\right)
_{\omega}dT+\left(  \frac{\partial\mathcal{U}}{\partial\omega}\right)
_{T}d\omega-\frac{\mathcal{U}}{\omega}d\omega
\end{equation}
Since the variables $\omega$ and $T$ are independent, this requires
\begin{equation}
T\left(  \frac{\partial\mathcal{S}}{\partial T}\right)  _{\omega}=\left(
\frac{\partial\mathcal{U}}{\partial T}\right)  _{\omega},~~~~~T\left(
\frac{\partial\mathcal{S}}{\partial\omega}\right)  _{T}=\left(  \frac
{\partial\mathcal{U}}{\partial\omega}\right)  _{T}-\frac{\mathcal{U}}{\omega}%
\end{equation}
Differentiating the first equation of (7) with respect to $\omega$ and the
second with respect to $T$ gives
\begin{equation}
T\frac{\partial^{2}\mathcal{S}}{\partial\omega\partial T}=\frac{\partial
^{2}\mathcal{U}}{\partial\omega\partial T},~~~~~\left(  \frac{\partial
\mathcal{S}}{\partial\omega}\right)  _{T}+T\frac{\partial^{2}\mathcal{S}%
}{\partial T\partial\omega}=\frac{\partial^{2}\mathcal{U}}{\partial
T\partial\omega}-\frac{1}{\omega}\left(  \frac{\partial\mathcal{U}}{\partial
T}\right)  _{\omega}%
\end{equation}
Subtracting the first equation of (8) from the second as as to eliminate the
second-derivative terms, we have
\begin{equation}
\left(  \frac{\partial\mathcal{S}}{\partial\omega}\right)  _{T}=-\frac
{1}{\omega}\left(  \frac{\partial\mathcal{U}}{\partial T}\right)  _{\omega}%
\end{equation}
Next using (9) together with the first equation of (7), we find
\begin{equation}
\left(  \frac{\partial\mathcal{S}}{\partial\omega}\right)  _{T}=-\frac
{T}{\omega}\left(  \frac{\partial\mathcal{S}}{\partial T}\right)  _{\omega}%
\end{equation}
which has the general solution $\mathcal{S}(T,\omega)=g(\omega/T)$ where $g$
is an arbitrary function of the single variable $\omega/T$. On the other hand
if we use the second equation of (7), then we find from Eq.(9)
\begin{equation}
\left(  \frac{\partial\mathcal{U}}{\partial\omega}\right)  _{T}-\frac
{\mathcal{U}}{\omega}=-\frac{T}{\omega}\left(  \frac{\partial\mathcal{U}%
}{\partial T}\right)  _{\omega}%
\end{equation}
which has the general solution
\begin{equation}
\mathcal{U}(T,\omega)=\omega f(\omega/T)
\end{equation}
where $f$ is an arbitrary function of the single variable $\omega/T$. This is
corresponds to the information in the Wien displacement theorem.[3] Although
the Wien theorem is often derived from the adiabatic compression of waves in a
cavity, our analysis shows that it holds in thermal equilibrium for any system
described by a simple harmonic oscillator Hamiltonian.

\noindent\textbf{THERMODYNAMIC FUNCTIONS FOR A HARMONIC OSCILLATOR}

The Wien displacement results\ $\mathcal{S}(T,\omega)=g(\omega/T)$ and
$\mathcal{U}(T,\omega)=\omega f(\omega/T)$ give constraints upon all the
thermodynamic functions for a harmonic oscillator. The thermodynamic
potential[4] $\phi(\omega/T)$, from which all the other thermodynamics
functions may be derived, must be a function of the combination $\omega/T$.
The average oscillator energy $\mathcal{U}$ in thermal equilibrium follows as
\begin{equation}
\mathcal{U}(T,\omega)=T^{2}\left(  \frac{\partial\phi}{\partial T}\right)
_{\omega}=-\omega\phi^{\prime}(\omega/T)
\end{equation}
The entropy $\mathcal{S}$ of the oscillator is again a function of $\omega
/T$,
\begin{equation}
\mathcal{S}(\omega/T)=\phi(\omega/T)+\mathcal{U}(T,\omega)/T=\phi
(\omega/T)-(\omega/T)\phi^{\prime}(\omega/T)
\end{equation}
The Helmholtz free energy $\mathcal{F}$ is directly related to the
thermodynamic potential $\phi(\omega/T)$
\begin{equation}
\mathcal{F}(T,\omega)=-T\phi(\omega/T)
\end{equation}
The generalized force $\mathcal{X}$ associated with a change in $\omega$ is
\begin{equation}
\mathcal{X}(\omega/T)=T\left(  \frac{\partial\phi}{\partial\omega}\right)
_{T}=\phi^{\prime}(\omega/T)
\end{equation}
and the specific heat $\mathcal{C}$ at constant $\omega$ is given by
\begin{equation}
\mathcal{C}(\omega/T)=\left(  \frac{\partial\mathcal{U}}{\partial T}\right)
_{\omega}=\left(  \frac{\omega}{T}\right)  ^{2}\phi^{\prime\prime}(\omega/T)
\end{equation}
Thus the equilibrium thermodynamics of a classical harmonic oscillator system
is determined by one function, the unknown function $\phi(\omega/T)$.

\noindent\textbf{WIEN DISPLACEMENT RESULT AND ZERO-POINT ENERGY}

There are two natural extremes for the oscillator energy given by the Wien
displacement result in (13); one extreme makes the energy $\mathcal{U}%
(T,\omega) $ independent of temperature $T$, and the other makes the energy
$\mathcal{U}(T,\omega)$ independent of the natural frequency $\omega$.

We deal first with the temperature-independent energy. When the potential
function $\phi^{\prime}(\omega/T)=const$ and so $\phi(\omega/T)=const(\omega
/T)$, then the oscillator energy in (13) takes the form
\begin{equation}
\mathcal{U}(T,\omega)=\mathcal{U}_{zp}(\omega)=const\times\omega
\end{equation}
This corresponds to temperature-independent zero-point energy.

We note that substitution of the zero-point energy (18) into the first law of
thermodynamics in the form
\begin{equation}
dQ=d\mathcal{U}-(\mathcal{U}/\omega)d\omega=d(const\times\omega)-(const\times
\omega/\omega)d\omega=0
\end{equation}
tells us that no heat $dQ$ enters the system on changing the natural frequency
of the oscillator $\omega$. Thus changes in zero-point energy occur without
any change in the thermodynamic entropy $\mathcal{S}(\omega/T)$ of the system.
Indeed, we see that if $\phi_{zp}^{\prime}$ is constant, then $\phi
_{zp}(\omega/T)=const\times\omega/T$ must be linear in its argument, and the
entropy $\mathcal{S}$ in Eq.(14) vanishes for any function $\phi$ which is
linear in its argument.

Classical zero-point energy is random energy which is present even at zero
temperature. Classical thermodynamics allows the possibility of classical
zero-point energy and experimental evidence (such as that for van der Waals
forces) requires its existence.[5] It is natural to choose the unknown
constant for the zero-point energy so as to fit the experimentally measured
van de Waals forces. This corresponds to an oscillator energy
\begin{equation}
\mathcal{U}_{zp}(\omega)=(1/2)\hbar\omega
\end{equation}
where $\hbar$ is a constant which takes the value familiar for Planck's constant.

\noindent\textbf{WIEN DISPLACEMENT RESULT AND ENERGY EQUIPARTITION}

The other extreme for the Wien displacement result (13) is the case where the
oscillator energy depends upon the temperature but has no dependence upon the
natural oscillator frequency $\omega$. Thus when $\phi^{\prime}(\omega
/T)=const/(\omega/T)$ in equation (13), then the oscillator energy is
\begin{equation}
\mathcal{U}(T,\omega)=\mathcal{U}_{RJ}(T)=const\times T.
\end{equation}
This is the familiar energy equipartition law (proposed by Rayleigh and Jeans
for low-frequency radiation modes) where the constant is chosen as Boltzmann's
constant $k_{B}$,
\begin{equation}
\mathcal{U}_{RJ}(T)=k_{B}T.
\end{equation}

In this case, an isothermal change of the natural oscillator frequency
$\omega$ produces no change in the oscillator internal energy. Rather, from
(5), the isothermal work done on changing the natural frequency $\omega$ is
provided by heat added which keeps the internal oscillator energy constant,
\begin{equation}
dQ=Td\mathcal{S}(\omega/T)=d\mathcal{U}_{RJ}(T)-(\mathcal{U}/\omega
)d\omega,~~~constant\text{ }T
\end{equation}
Then
\begin{equation}
d\mathcal{S}_{RJ}=0-(k_{B}/\omega)d\omega,~~~constant\text{ }T,
\end{equation}
and since we know the functional form $\mathcal{S}(\omega/T)$, we have the
familiar result
\begin{equation}
\mathcal{S}_{RJ}(\omega/T)=-k_{B}\ln(\omega/T)+const
\end{equation}
Indeed if $\phi_{RJ}^{\prime}(\omega/T)=-k_{B}/(\omega/T)$, then $\phi
_{RJ}(\omega/T)=-\ln(\omega/T)$ and the entropy in (14) takes the form (25).

\noindent\textbf{USE OF NATURAL UNITS IN THE ANALYSIS}

In this paper we are not interested in the numerical evaluation of
thermodynamic quantities but rather in the fundamental connections between
thermal energy and zero-point energy. On this account we will measure all
quantities in terms of energy and take the entropy as a pure number. Thus we
will take $\hbar=1$ and measure frequencies in energy units. Also, we will
take $k_{B}=1$ and measure temperature in energy units. Finally, we note that
we can choose to take the speed of light $c=1$ and measure distances in
inverse energy units.[6] \ Thus the limiting form corresponding to zero-point
energy has
\begin{equation}
\phi_{zp}(\omega/T)=-(1/2)(\omega/T),~~~~~\mathcal{U}_{zp}(\omega)=(1/2)\omega
\end{equation}
while the limiting form corresponding to energy equipartition takes the form
\begin{equation}
\phi_{RJ}(\omega/T)=-\ln(\omega/T),~~~~~\mathcal{U}_{RJ}(T)=T
\end{equation}

\noindent\textbf{ASYMPTOTIC LIMITS FOR THERMAL OSCILLATOR ENERGY}

In general, the energy of an oscillator system will depend upon both frequency
$\omega$ and temperature $T$ as in (13). In the limit as $T\rightarrow0$, we
expect to recover the zero-point energy of the oscillator
\begin{equation}
lim_{T\rightarrow0}\mathcal{U}(T,\omega)=lim_{T\rightarrow0}[-\omega
\phi^{\prime}(\omega/T)]=\mathcal{U}_{zp}(\omega)=(1/2)\omega
\end{equation}
and in the limit $\omega\rightarrow0$, we expect to obtain the equipartition
energy
\begin{equation}
lim_{\omega\rightarrow0}\mathcal{U}(T,\omega)=lim_{\omega\rightarrow0}%
[-\omega\phi^{\prime}(\omega/T)]=\mathcal{U}_{RJ}(T)=T
\end{equation}

It is useful to make a distinction between the THERMAL energy $\mathcal{U}%
_{T}(T,\omega)$ of an oscillator and the oscillator's TOTAL energy
$\mathcal{U}(T,\omega)$. The thermal energy is just the (average) energy above
the (average) zero-point energy
\begin{equation}
\mathcal{U}_{T}(T,\omega)=\mathcal{U}(T,\omega)-\mathcal{U}_{zp}%
(\omega)=-\omega\phi^{\prime}(\omega/T)-(1/2)\omega=-\omega\lbrack\phi
^{\prime}(\omega/T)-\phi_{zp}^{\prime}(\omega/T)]
\end{equation}
Although the total oscillator energy $\mathcal{U}(T,\omega)$ is related to
forces, it is only the thermal oscillator energy $\mathcal{U}_{T}(T,\omega)$
which is related to changes in thermodynamic entropy since (as seen above)
$\phi_{zp}(\omega/T)$ does not give any thermodynamic entropy. \ 

\noindent\textbf{THERMAL OSCILLATOR ENERGY AS THE SMOOTHEST INTERPOLATION
BETWEEN THE EQUIPARTITION AND ZERO-POINT LIMITS}

The two limiting forms (26) and (27) which make the average energy independent
of one of the two variables $T$ or $\omega$ correspond to the expected
limiting cases of high freqency-low temperature $\omega/T>>1$ or low
frequency-high temperature $\omega/T<<1$ in (28) and (29). It is clear that
the actual thermodynamic functions for a classical harmonic oscillator should
all be smooth functions between these asymptotic forms. Prompted by our ideas
that thermodynamic functions represent the maximum of some quantity in the
presence of constraints, we suggest that "the equilibrium thermodynamic
behavior of a harmonic oscillator is given by that function satisfying the
Wien displacement results which provides the smoothest possible interpolation
between the asymptotic forms given by energy equipartition at low frequency
and zero-point energy at high frequency." Let us explore this idea.

The limiting thermodynamic potentials
\begin{equation}
\phi_{RJ}(\omega/T)=-\ln(\omega/T),~~~~~\phi_{zp}(\omega/T)=-(1/2)\omega/T
\end{equation}
involve a logarithmic behavior at the low-frequency limit. Since the
logarithmic function is more complicated analytically than the exponential
function, it is convenient to take the exponential of the negative of these
functions and to consider
\begin{equation}
\exp[-\phi_{RJ}(\omega/T)]=\omega/T,~~~\exp[-\phi_{zp}(\omega/T)]=exp[\omega
/(2T)]
\end{equation}
The exponentiation will not change the "smoothest possible" criterion required
of the interpolation. \ Thus we are searching for the "smoothest possible
interpolation" between linear behavior $\omega/T$ at small argument and
exponential behavior $exp[\omega/(2T)]$\ at large argument. A familiar
analytic function which joins exactly these asymptotic limits is the
hyperbolic sine function. Thus we write this smooth interpolation function as
\begin{equation}
\exp\left[  -\phi_{Pzp}\left(  \frac{\omega}{T}\right)  \right]
=2\sinh\left(  \frac{1}{2}\frac{\omega}{T}\right)
\end{equation}
It is easy to see that the right-hand side of (33) has exactly the asymptotic
forms demanded in (32). This smooth interpolation (33) then leads to the
thermodynamic functions
\begin{equation}
\phi_{Pzp}\left(  \frac{\omega}{T}\right)  =-\ln\left[  2\sinh\left(  \frac
{1}{2}\frac{\omega}{T}\right)  \right]
\end{equation}%
\begin{equation}
\mathcal{U}_{Pzp}\left(  T,\omega\right)  =\frac{1}{2}\omega\coth\left(
\frac{1}{2}\frac{\omega}{T}\right)  =\frac{\omega}{\exp(\omega/T)-1}+\frac
{1}{2}\omega
\end{equation}%
\begin{equation}
\mathcal{S}_{P}\left(  \frac{\omega}{T}\right)  =-\ln\left[  2\sinh\left(
\frac{1}{2}\frac{\omega}{T}\right)  \right]  +\frac{1}{2}\frac{\omega}{T}%
\coth\left(  \frac{1}{2}\frac{\omega}{T}\right)
\end{equation}
We have labeled these thermodynamic functions with the subscript "Pzp" or "P"
because they corresponds exactly to those related to the Planck average
oscillator energy including zero-point energy. \ As noted above, the entropy
depends upon the Planck thermal spectrum but does not reflect the zero-point energy.

Thus demanding the smoothest interpolation between the equipartition and
zero-point limits suggests the Planck spectrum. Certainly, the hyperbolic sine
function is smooth along the real axis in the sense that it is an analytic
function, and the function, together with all its derivatives, is monotonic
for positive argument. Also, the differences between the hyperbolic sine
function (33) and both asymptotic forms in (32) are monotonic functions and
indeed all the derivatives of the differences are monotonic functions for
positive argument. Clearly a "smooth" interpolation should lead to
thermodynamic functions all of which are monotonic (demanded by the
thermodynamics) and which have monotonic differences with the asymptotic limit
functions. The Planck result with zero-point energy in Eq.(34) indeed leads to
"smooth" interpolations for all the thermodynamic functions. Because of the
monotonic behavior for the thermodynamic potential which includes higher
derivatives, we suggest that there is a natural sense in which the Planck
result is the "smoothest possible" interpolation between the equipartition and
zero-point energy limits. However, someone more mathematically knowledgeable
than the author must judge whether the "smoothness" criterion can easily be
made numerical.

\noindent\textbf{PART II: THERMODYNAMICS OF WAVES IN A BOX}

\noindent\textbf{USE OF ONE DIMENSIONAL WAVES FOR SIMPLICITY}

Because we have not been able to give a numerical criterion for "the smoothest
possible interpolation between the equipartition and zero-point asymptotic
forms," we will turn to Casimir energies within the full radiation problem
where such a criterion seems easily possible. Once again we will find that
thermal radiation is intimately connected with zero-point radiation.

Although the calculations to be described below can be carried through for
electromagnetic waves in a three-dimensional box, we will consider a
thermodynamic wave system in one spatial dimension rather than in three
because the mathematics is distinctly simpler while the physical ideas are
unchanged.[7] Thus we can imagine one-dimensional thermodynamic systems
consisting of beads on parallel frictionless wires which move between two
walls, or of waves on a string, or of electromagnetic waves which are required
to move between two conducting walls with wave vectors $\mathbf{{k}}$ which
are always perpendicular to the walls. For electromagnetic waves in this case
of one spatial dimension, the pressure $p$ on the walls would be related to
the energy density $\mathcal{E}/V$ as
\begin{equation}
p=\mathcal{E}/V
\end{equation}
which holds for normally incident plane waves, rather than the
three-dimensional situation $p=(1/3)\mathcal{E}/V$ which holds for waves
incident on a wall when averaged over all available directions. If we multiply
by the area of the walls, then Eq.(37) becomes
\begin{equation}
X=u
\end{equation}
for our waves in one spatial dimension where $X$ is the force on a bounding
partition and $u$ is the energy per unit length. [8]

\noindent\textbf{DERIVATION OF THE STEFAN-BOLTZMANN LAW}

For waves in one spatial dimension, we can use the usual thermodynamic
arguments to obtain the corresponding Stefan-Boltzmann law for the thermal
radiation energy which is connected to thermodynamic entropy. Thus we assume
that at thermal equilibrium the thermal energy $U_{T}(T,L)$ and entropy
$S(T,L)$ of our waves in a very large one-dimensional box of length $L$
satisfy
\begin{equation}
U_{T}(T,L)=Lu_{T}(T),~~~~~S(T,L)=Ls(T)
\end{equation}
so that, in this large-L limit, the thermal energy density $u_{T}$ and entropy
density $s$ are functions of temperature only. Then the laws of thermodynamics
give
\begin{equation}
TdS(T,L)=dU_{T}(T,L)+X_{T}dL
\end{equation}
where we connect the thermal-related force $X_{T}$ to the thermal-related
energy density $u_{T}$, $X_{T}=u_{T}$ as in (38). Then from (39) and (40),
\begin{equation}
T\left(  sdL+L\frac{ds}{dT}dT\right)  =u_{T}dL+L\frac{du_{T}}{dT}dT+u_{T}dL
\end{equation}
Comparing differentials on both sides, we have
\begin{equation}
s=\frac{2u_{T}}{T},~~~~~\frac{ds}{dT}=\frac{1}{T}\frac{du_{T}}{dT}%
\end{equation}
Differentiating the first equation in (42) with respect to temperature and
substituting into the second, we find
\begin{equation}
\frac{-2u_{T}}{T^{2}}+\frac{2}{T}\frac{du_{T}}{dT}=\frac{1}{T}\frac{du_{T}%
}{dT}%
\end{equation}
with solution
\begin{equation}
u_{T}=\alpha T^{2}%
\end{equation}
where $\alpha$ is an unknown constant. This corresponds to the
Stefan-Boltzmann law for this one-spatial-dimension system. Also the entropy
per unit length $s$ follows from (42) and (44) as
\begin{equation}
s=2\alpha T
\end{equation}
Had we used the 3-dimensional form $p=(1/3)u$ and the same procedure with
volume $V$ replacing the length $L$, we would have found the familiar
expressions $u_{T}=\alpha T^{4}$ and $s=(4/3)\alpha T^{3}$.[9]

\noindent\textbf{WAVE NORMAL MODES IN A ONE-DIMENSIONAL BOX}

Systems satisfying the wave equation in a container with conducting walls can
be described in terms of normal modes of oscillation, each of which
corresponds to a harmonic oscillator system[10]
\begin{equation}
L(q_{\lambda},\dot{q}_{\lambda})=\Sigma_{\lambda}(1/2)(\dot{q}_{\lambda}%
^{2}-\omega_{n}^{2}q_{\lambda}^{2})
\end{equation}
where the $q_{\lambda}$ are the amplitudes of the normal modes. For waves in
one spatial dimension inside a box of length $L$, the normal modes can be
labeled by a single integer index $n$ where the associated frequency
$\omega_{n}$ is given by $\omega_{n}=cn\pi/L$, $n=1,2,3...$, where $c$ is the
speed of the waves. For light, we can choose $c=1$. Accordingly we can use the
harmonic-oscillator Wien-displacement result (12) or (13) for each normal mode
to obtain the Stefan-Boltzmann law for the thermal part of $\mathcal{U}_{T}$
of the radiation energy. We consider only the thermal energy $\mathcal{U}_{T}$
since this is a finite quantity when summed over all normal modes, in contrast
to the total energy $\mathcal{U}$ or zero-point energy $\mathcal{U}_{zp}$ both
of which diverge in the sum over (infinitely many) high-frequency modes. In a
one-dimensional box which is so large that the discrete sum over normal modes
can be replaced by an integral, we have the total thermal energy $U_{T}(T,L)$
given by
\[
U_{T}(T,L)=\Sigma_{n=1}^{\infty}\mathcal{U}_{T}(T,\frac{n\pi c}{L}%
)=\Sigma_{n=1}^{\infty}\frac{n\pi c}{L}\left[  \phi^{\prime}(\frac{n\pi c}%
{LT})-\frac{1}{2}\right]
\]

\begin{equation}
\approx\int_{0}^{\infty}dn\frac{n\pi c}{L}\left[  \phi^{\prime}(\frac{n\pi
c}{LT})-\frac{1}{2}\right]  =\frac{L}{c\pi}T^{2}\int_{0}^{\infty}dzz\left[
\phi^{\prime}(z)-\frac{1}{2}\right]
\end{equation}
for one space dimension. This is just the Stefan-Boltzmann result obtained
earlier in Eq.(44). In the case of 3 spatial dimensions, the frequencies of
the normal modes $\omega_{lmn}$ would be labeled by 3 integer indices and the
same procedure would lead to a $T^{4}$ temperature dependence for a large container.

\noindent\textbf{THERMODYNAMICS OF CASIMIR FORCES}

The Stefan-Boltzmann law given in (44) and (47) connects the total thermal
radiation energy per unit length to the thermodynamic temperature $T$, and
(through Eq.(38)) gives the thermal forces for a large box. However, it
provides no information regarding the spectrum of thermal radiation. Now in
obtaining Eq.(47), we took the limit of a large box $L$ and so replaced the
sum over normal modes by an integral. However, by going to the continuum
limit, we lost the information which might be available in the discrete
spectrum of the normal modes. It was Casimir who saw the possibility of new
forces and energies linked to this discreteness of the classical normal mode
structure. The most famous example of such forces is the original Casimir
calculation[11] of the force between conducting parallel plates arising from
electromagnetic zero-point radiation. Casimir worked specifically with
zero-point fields; however, the idea is not limited to zero-point radiation.
Any spectrum of random classical radiation will lead to Casimir energies
associated with the discrete classical normal mode structure of a container.
Indeed, every thermodynamic variable will depend upon the normal modes structure.

It should be emphasized how totally different this classical wave situation is
from the classical particle situation of ideal gas particles in a box. Thus if
a box with reflecting walls is filled with ideal gas particles at temperature
$T$, then the introduction of a thin reflecting partition does not change the
system energy and does not involved any average force on the partition. In
total contrast, the introduction of a conducting partition into a
conducting-walled box of thermal radiation leads to a change in the normal
mode structure and hence both to position-dependent energy changes (Casimir
energies) and to average forces on the partition (Casimir forces). These
Casimir energies and forces will depend upon the precise spectrum of random
radiation and upon the precise location of the partition. In this article we
are proposing that the Planck spectrum for thermal radiation equilibrium
arises from a natural minimum principle for the Casimir energy changes
associated with the placement of a partition in a box of radiation.

\noindent\textbf{CHANGE IN CASIMIR ENERGY DUE TO A PARTITION}

We now consider a one-dimensional box of length $L$ and calculate the change
of radiation energy $\Delta U(x,L,T)$ with position $x$ for a partition which
is located a distance $x$ from the left-hand end of the box, $0\leq x\leq L$.
The energy of each normal mode of frequency $\omega_{n}$ is given by
$\mathcal{U}(T,\omega_{n})$. The partition changes the normal mode frequencies
and so produces a position-dependent energy change $\Delta U(x,L,T)$ which is
a Casimir energy. We will calculate the Casimir energy $\Delta U(x,L,T)$ as
the change in the system energy when the partition is placed a distance $x$
from the left-hand wall compared to when the partition is placed at $x=L/2$ in
the middle of the box,
\[
\Delta U(x,L,T)=\left\{  U(T,x)+U(T,L-x)\right\}  -\left\{
U(T,L/2)+U(T,L/2)\right\}
\]%
\begin{equation}
=\left\{  \Sigma_{n=1}^{\infty}\mathcal{U}\left(  T,\frac{cn\pi}{x}\right)
+\Sigma_{n=1}^{\infty}\mathcal{U}\left(  T,\frac{cn\pi}{L-x}\right)  \right\}
-2\Sigma_{n=1}^{\infty}\mathcal{U}\left(  T,\frac{cn\pi}{L/2}\right)
\end{equation}

\noindent\textbf{CASIMIR ENERGY FOR THE ZERO-POINT SPECTRUM}

We consider first the two extreme cases for the thermodynamic potential
$\phi(\omega/T)$ which make the energy spectrum $\mathcal{U}\left(
T,\omega\right)  $ independent of one of its two variables. The zero-point
spectrum (26) and the Rayleigh-Jeans spectrum (27) both give divergent
energies when summed over the infinite number of normal modes. However, this
need not deter us from calculating the Casimir energies associated with these
spectra. The Casimir energies $\Delta U_{zp}(x,L)$ and $\Delta U_{RJ}(x,L,T)$
are defined as limits and are finite. In contrast to an ideal system, any
physical system (such as a string with clamped ends or else electromagnetic
fields in a region bounded by good conductors) will not enforce the normal
mode structure at very high frequencies (short wavelengths). Thus it is
natural to introduce a smooth cut-off $\exp(-\Lambda\omega/c)$ related to
frequency $\omega=ck$
\begin{equation}
U(T,L,\Lambda)=\Sigma_{n=1}^{\infty}\mathcal{U}(T,\omega_{n})\exp
(-\Lambda\omega_{n}/c)
\end{equation}
Next we carry out the subtractions corresponding to (48) to obtain the Casimir
energy, $\Delta U(x,L,T,\Lambda)$, and then allow the no-cut-off limit
$\Lambda\rightarrow0$. Although here we will work with an exponential cut-off
because it is easy to sum the geometric series, the result is very general;
any smooth cut-off function dependent on frequency alone will give the same result[12].

In this fashion, we obtain the Casimir energy for the zero-point radiation
spectrum (26),
\[
\Delta U_{zp}(x,L)=lim_{\Lambda\rightarrow0}\left\{  \Sigma_{n=1}^{\infty
}\frac{1}{2}\frac{cn\pi}{x}\exp\left(  -\Lambda\frac{n\pi}{x}\right)
+\right.
\]%
\[
\left.  +\Sigma_{n=1}^{\infty}\frac{1}{2}\frac{cn\pi}{L-x}\exp\left(
-\Lambda\frac{n\pi}{L-x}\right)  -2\Sigma_{n=1}^{\infty}\frac{1}{2}\frac
{cn\pi}{L/2}\exp\left(  -\Lambda\frac{n\pi}{L/2}\right)  \right\}
\]%
\[
=lim_{\Lambda\rightarrow0}\left\{  -\frac{c}{2}\frac{\partial}{\partial
\Lambda}\left[  \frac{1}{\exp(\frac{\Lambda\pi}{x})-1}+\frac{1}{\exp
(\frac{\Lambda\pi}{L-x})-1}-2\frac{1}{\exp(\frac{\Lambda\pi}{L/2})-1}\right]
\right\}
\]%
\[
=lim_{\Lambda\rightarrow0}\left\{  \left[  \frac{cx}{2\Lambda^{2}\pi}%
-\frac{c\pi}{24x}+\bigcirc(\Lambda)\right]  +\left[  \frac{c(L-x)}%
{2\Lambda^{2}\pi}-\frac{c\pi}{24(L-x)}+\bigcirc(\Lambda)\right]  -\right.
\]%
\begin{equation}
\left.  -2\left[  \frac{c(L/2)}{2\Lambda^{2}\pi}-\frac{c\pi}{24(L/2)}%
+\bigcirc(\Lambda)\right]  \right\}  =-\frac{c\pi}{24}\left(  \frac{1}%
{x}+\frac{1}{L-x}-\frac{2}{L/2}\right)
\end{equation}
Thus we obtain the change in zero-point energy associated with the position
$x$ of the partition,
\begin{equation}
\Delta U_{zp}(x,L)=-\frac{c\pi}{24}\left(  \frac{1}{x}+\frac{1}{L-x}-\frac
{2}{L/2}\right)
\end{equation}

\noindent\textbf{CASIMIR ENERGY FOR THE RAYLEIGH-JEANS SPECTRUM}

In an analogous calculation we obtain the Casimir energy for the
Rayleigh-Jeans spectrum (27),
\[
\Delta U_{RJ}(x,L,T)=
\]%
\[
=lim_{\Lambda\rightarrow0}\left\{  \Sigma_{n=1}^{\infty}T\exp\left(
-\Lambda\frac{n\pi}{x}\right)  +\Sigma_{n=1}^{\infty}T\exp\left(
-\Lambda\frac{n\pi}{L-x}\right)  -2\Sigma_{n=1}^{\infty}T\exp\left(
-\Lambda\frac{n\pi}{L/2}\right)  \right\}
\]%
\[
=lim_{\Lambda\rightarrow0}\left\{  \frac{T}{\exp(\frac{\Lambda\pi}{x}%
)-1}+\frac{T}{\exp(\frac{\Lambda\pi}{L-x})-1}-2\frac{T}{\exp(\frac{\Lambda\pi
}{L/2})-1}\right\}
\]%
\[
=lim_{\Lambda\rightarrow0}\left\{  T\left[  \frac{x}{\Lambda\pi}-\frac{1}%
{2}+\frac{\pi\Lambda}{12x}-...\right]  +T\left[  \frac{L-x}{\Lambda\pi}%
-\frac{1}{2}+\frac{\pi\Lambda}{12(L-x)}-...\right]  +\right.
\]%
\begin{equation}
\left.  -2T\left[  \frac{L/2}{\Lambda\pi}-\frac{1}{2}+\frac{\pi\Lambda
}{12(L/2)}-...\right]  \right\}  =0
\end{equation}
The Rayleigh-Jeans spectrum is the unique spectrum which produces no Casimir
energy changes associated with the placement of the Casimir partition, $\Delta
U_{RJ}(x,L,T)=0$.

\noindent\textbf{AN EXTREMUM PRINCIPLE FOR THERMAL RADIATION}

We note the striking difference in the Casimir energy for these two limiting
cases. The Rayleigh-Jeans spectrum gives $\Delta U_{RJ}(x,L,T)=0$,
corresponding to complete independence of the Casimir energy from the location
$x$ of the partition. This is just like the ideal-gas particle situation where
the system energy is independent of the partition location. In contrast, the
zero-point spectrum involves a smooth change in the Casimir energy as given in
Eq.(51) with a divergent change as the separation $x$ from the wall decreases
to zero; the Casimir spectrum is special because it involves no entropy change
with the location of the partition.

Just as we suggested earlier that the thermodynamic behavior of a harmonic
oscillator system could be obtained as the smoothest possible transition from
the equipartition limit to the zero-point limit, here we suggest the same idea
for the Casimir energies arising from thermal radiation. We expect that the
thermal radiation spectral function represents the smoothest possible
transition between the two extreme cases for Casimir energy changes calculated
above. Thus, for fixed temperature $T$, we want to find that spectrum
$\mathcal{U}(T,\omega)=-\omega\phi^{\prime}(\omega/T)$ which is a monotonic
interpolation between energy equipartition at temperature $T$ at low frequency
and zero-point energy $(1/2)\omega$ at high frequency which makes the Casimir
energy $\Delta U(x,L,T)$ depart from zero by the least possible amount. In
other words, the position of the partition plays as little role as possible in
changing the system energy. This assumption arises from our sense that the
thermal radiation spectrum should give the greatest possible energy uniformity
consistent with finite thermal energy. We recall that for ideal gas particles
at thermal equilibrium, the system energy has no dependence at all on the
partition's location.

We must turn this qualitative suggestion into a quantitative criterion. Since
the Casimir energy vanishes at the middle of the box, $\Delta U(L/2,L,T)=0$,
any departure of the Casimir energy from zero represents a failure of
uniformity. Specifically, at fixed finite temperature $T$, let us minimize the
integral $I$ of the absolute value of the Casimir energy integrated over the
length of the box
\begin{equation}
I=\int_{x=\delta}^{x=R/2}dx~|\Delta U(x,L,T)|
\end{equation}
where $\delta$ is a small cut-off distance which is much less than any other
length in the situation, $0<\delta<<min(L,c/T)$. The need for such a cut-off
arises because of the divergence of the zero-point Casimir energy at small distances.

\noindent\textbf{CASIMIR ENERGIES FOR VARIOUS RADIATION SPECTRA}

In one spatial dimension, it is quick to evaluate the Casimir energies for
various monotonic spectral functions $\mathcal{U}(T,\omega)=\omega
f(\omega/T)$ on a home computer. One separates out the divergent zero-point
energy contribution corresponding to (51) and then evaluates the thermal
contribution to the Casimir energy for any assumed thermal spectrum
$\mathcal{U}_{T}(T,\omega)=Tg(\omega/T)$ where $g(0)=1$, corresponding to the
equipartition limit at low frequency, and also $\int_{0}^{\infty}%
dzg(z)<\infty$, corresponding to finite thermal radiation density in space.
The total Casimir energy $\Delta U(x,L,T)$ is the sum of the thermal and
zero-point contributions. \ 

We require that $\mathcal{U}(T,\omega)$ be a monotonically increasing function
of frequency as it goes from the low frequency limit $\mathcal{U}(T,\omega)=T$
to the high-frequency limit $\mathcal{U}(T,\omega)=(1/2)\omega$. \ This
already puts restrictions on the monotonically decreasing functions
$g(\omega/T)$. However, we find that even monotonically decreasing functions
$g(\omega/T)$ which satisfy the limits $g(0)=1$ and $\int_{0}^{\infty
}dzg(z)<\infty$ as well as giving monotonic functions $\mathcal{U}%
(T,\omega)=Tg(\omega/T)+(1/2)\omega$ still do not necessarily lead to
monotonic Casimir energy changes $\Delta U(x,L,T)$ with partition position
$x$. \ Such functions are excluded by fundamental thermodynamics ideas.
\ Nevertheless, for all spectra $\mathcal{U}_{T}(T,\omega)$ satisfying the two
required limiting conditions, we can calculate the test integral given in
Eq.(53). The spectrum providing the smallest value for the integral appears to
be the Planck spectrum with zero-point radiation. Indeed, we may use the
Planck form (or other functional forms) in a variational calculation to obtain
the parameters which give the smallest value to the test integral for the
given functional form. Thus the Planck form for the thermal part of the
radiation at frequency $\omega$ can be written as
\begin{equation}
\mathcal{U}_{PT}(T,\omega)=\mathcal{U}_{P}(T,\omega)-\frac{1}{2}\omega
=\frac{\omega}{\exp(\omega/T)-1}%
\end{equation}
We can introduce parameters $C_{1}$ and $C_{2}$ into a generalization of this
form giving an energy spectrum including zero-point energy as
\begin{equation}
\mathcal{U}_{C_{1}C_{2}}(T,\omega)=\frac{C_{1}\omega\exp[-C_{2}(\omega
/T)]}{1-\exp[-C_{1}(\omega/T)]}+(1/2)\omega
\end{equation}
For all positive parameters $C_{1}$ and $C_{2}$, this spectrum goes over to
energy equipartition in the limit $\omega\rightarrow0$ and goes over to
zero-point energy at high frequency while giving finite total thermal
radiation energy. Accordingly we can search for the the values of $C_{1}$ and
$C_{2}$ which make the test integral (53) a minimum. Numerical calculation
shows that the minimum value for the test integral is achieved when
$C_{1}=C_{2}=1$, corresponding to exactly the Planck spectrum (35).

Indeed, of all the functional forms tested numerically, the Planck spectrum
gave the smallest value of the test integral. We conjecture that analytic
calculation would show that this spectrum provides the minimum for this
integral, and hence in this sense provides the smallest Casimir energies in
the presence of zero-point radiation.[13]

\noindent\textbf{'ULTRA-VIOLET CATASTROPHE' WITHOUT ZERO-POINT RADIATION}

We should also note that our minimum principle indeed requires the presence of
zero-point energy. If no zero-point energy were present, then we would still
require that the thermal spectrum give energy equipartition at low frequency
and go to zero at high frequency so as to give a finite energy density for
thermal radiation. For this case, the thermal energy would be the total energy
in (53). \ However, there would be no natural high-frequency limit. If we
tried a smooth spectrum such as that suggested by Rayleigh $\mathcal{U}%
_{RT}(T,\omega)=T\exp[-C(\omega/T)]$ with an adjustable parameter $C$ but
without zero-point energy, then we would find that the test integral given in
Eq.(53) decreases as the parameter $C$ decreases, bringing the spectrum ever
closer to the Rayleigh-Jeans spectrum, in which limit the integral vanishes
$I=0$ and there are no Casimir energy changes. The absence of any natural
cut-off frequency represents behavior reminiscent of the "ultraviolet
catastrophe" emphasized by Einstein and named by Ehrenfest in 1911. What
prevents the catastrophic shift of thermal radiation to ever-higher
frequencies is precisely the presence of zero-point radiation.

\noindent\textbf{ENTROPY CHANGES IN THE HIGH-TEMPERATURE LIMIT}

For fixed frequency $\omega$ and increasing temperature $T$, the Planck
spectrum with zero-point radiation goes over to the Rayleigh-Jeans spectrum.
Similarly, for fixed length $x$ and increasing temperature $T$ the Casimir
energy $\Delta U_{Pzp}(x,L,T)$ for the Planck spectrum with zero-point
radiation goes over to the Rayleigh-Jeans result $\Delta U_{RJ}(x,L,T)=0$.
However, the net force $\Delta X_{Pzp}(x,L,T)$\ and the net change in entropy
$\Delta S_{P}(x,L,T)$ do not go to zero with increasing temperature. \ Indeed,
these quantities go over to the values for the Rayleigh-Jeans spectrum, which
values do not vanish. Thus at high temperature, Casimir forces are associated
with changes in system entropy, not system energy.

The net force on the conducting partition in a box of radiation is calculated
by summing the forces due to the radiation in the normal modes on either side
of the partition. Since the total force on one side alone diverges with the
infinite number of normal modes, we introduce a frequency-dependent cut-off,
carry out the subtraction of the forces on the opposite sides, and then take
the no-cut-off limit. The force $X_{n}$ associated with a change in box length
$L$ and due to the $n$th mode is related to the force in (4) and (16)
$\mathcal{X}_{n}=$ $-\mathcal{U}_{n}/\omega_{n\text{ }}$associated with a
change in $\omega_{n}$ by%

\begin{equation}
X_{n}=\mathcal{X}_{n}(d\omega_{n}/dL)
\end{equation}
Then we obtain the net $\Delta X$ force on a conducting partition in our
one-dimensional conducting-walled box as a sum over the contributions from the
$X_{n}$ on both sides. \ For the Rayleigh-Jeans spectrum, this is%

\[
\Delta X_{RJ}(x,L,T)=lim_{\Lambda\rightarrow0}\left\{  \Sigma_{n=1}^{\infty
}\frac{-T}{cn\pi/x}\frac{-cn\pi}{x^{2}}\exp\left(  -\Lambda\frac{cn\pi}%
{x}\right)  +\right.
\]

\begin{equation}
\left.  +\Sigma_{n=1}^{\infty}\frac{-T}{cn\pi/(L-x)}\frac{cn\pi}{(L-x)^{2}%
}\exp\left(  -\Lambda\frac{cn\pi}{L-x}\right)  \right\}
\end{equation}
Once again we are dealing with a geometrical series which can be summed easily
to give%

\begin{equation}
\Delta X_{RJ}(T,x,L)=\frac{-T}{2}\left(  \frac{1}{x}-\frac{1}{L-x}\right)
\end{equation}
We notice that this force, which arises from the discrete classical normal
mode structure of the box, is significant only if $x$ or $L-x$ is small
compared to $T$.

Then applying thermodynamic analysis for a one-dimensional box of radiation at
a uniform temperature $T$ with a partition, an isothermal change of the the
position of the partition gives no change in system energy but involves heat
added so that%

\begin{equation}
Td\Delta S_{RJ}=d\Delta U_{RJ}+\Delta X_{RJ}dx=0+\Delta X_{RJ}dx
\end{equation}

\begin{equation}
d\Delta S_{RJ}=\frac{X_{RJ}}{T}dx=\frac{-1}{2}\left(  \frac{1}{x}-\frac
{1}{L-x}\right)  dx
\end{equation}

and%

\begin{equation}
\Delta S_{RJ}(x,L,T))=\frac{1}{2}\left\vert \ln\left(  \frac{x}{L-x}\right)
\right\vert
\end{equation}
We emphasize that the right-hand side of Eq.(61) is independent of temperature
$T$. Thus due to the normal mode structure of the box, there is a
temperature-independent change of entropy when the partition is moved. This
seems reminiscent of the temperature-independent change associated with mixing
entropy for ideal gas particles. This same result (61) can be obtained by
summing the entropy over the normal modes of the box when using a cut-off
function, subtracting, and allowing the cut-off to disappear. \ This sort of
entropy change was noted recently in quantum field theoretic analyses of
Casimir forces in three-dimensional situations involving electromagnetic radiation.[14]

The Casimir entropy change $\Delta S_{P}(x,L,T)=S_{P}(T,x)+S_{P}%
(T,L)-2S_{P}(T,L/2)$ associated with the Planck spectrum (and involving
convergent sums) goes to the Rayleigh-Jeans limit (61) at high temperature for
fixed values of $x$ and $L$. However, for fixed finite temperature $T$, the
entropy change $\Delta S_{P}(x,L,T)$ for the Planck spectrum goes to a finite
temperature-dependent limit as $x$ goes to $0$ or $L$.

\noindent\textbf{PART III: REVIEW OF CONNECTIONS BETWEEN THERMAL AND
ZERO-POINT ENERGIES}

\noindent\textbf{CLASSICAL ZERO-POINT RADIATION}

The discussion of the thermodynamics of blackbody radiation given here
involves two ideas which are usually absent from discussions of thermodynamics
in classical physics: i)classical zero-point radiation and ii)Casimir forces.
Both these ideas have appeared in the physics literature for some time but
have not yet been accepted into textbooks of classical physics.

Classical zero-point energy is random energy which is present even at zero
temperature. Classical thermodynamics allows the possibility of classical
zero-point energy and experimental evidence requires its existence. Thus
classical electromagnetic theory derived from Maxwell's differential equations
requires boundary conditions on these differential equations. Although
traditional classical electron theory[15] assumes vanishing electromagnetic
fields in the far past $E^{in}\rightarrow0$, $B^{in}\rightarrow0$, this
boundary condition is in contradiction to Nature. Experimental measurements of
van der Waals forces[5] require the existence of classical electromagnetic
zero-point radiation with a Lorentz-invariant spectrum as the boundary
conditions on Maxwell's equations. The unique Lorentz-invariant spectrum[16]
of random classical electromagnetic radiation corresponds to an energy
$\mathcal{U}_{zp}(\omega)=const\times\omega$ per normal mode. In order to fit
the experiments, the constant must be chosen as half Planck's constant,
$const=\hbar/2$.

Amongst physicists today, zero-point radiation is regarded as a "quantum"
phenomenon.[17] Actually however, zero-point radiation needs have nothing to
do with "quanta," with discontinuous in contrast to continuous energy changes.
Classical zero-point radiation is classical random radiation, just as thermal
radiation was regarded as classical random radiation in the years before 1900.
When the appropriate classical boundary conditions including zero-point
radiation are applied, classical physics can describe far more of Nature than
is credited in the current physics textbooks[18].

\noindent\textbf{PREVIOUS CLASSICAL DERIVATIONS OF THE PLANCK SPECTRUM}

Derivations of the blackbody radiation spectrum within classical physics have
taken many different forms, but all involve the presence of classical
zero-point radiation. In contrast to the one-dimensional thermodynamic
analysis given here, all the previous work involves classical electromagnetic
radiation in three spatial dimensions. The derivations include the following.
1) The original Einstein-Hopf calculation for the average translational
kinetic energy of a point electric dipole oscillator in a box with thermal
electromagnetic radiation does not include zero-point radiation; when
classical electromagnetic zero-point radiation is included, the Einstein-Hopf
analysis leads to the Planck spectrum with zero-point radiation as the thermal
radiation spectrum.[19] 2) The modification of the Einstein fluctuation
argument in the presence of classical zero-point radiation also gives the
Planck spectrum with zero-point radiation. \ Furthermore, we find an
understanding of photon-like fluctuations in terms of fluctuations related to
classical zero-point radiation.[20] 3) An analysis of the diamagnetic behavior
of a free particle in classical thermal radiation including zero-point
radiation leads to the Planck spectrum by comparison with a the behavior of a
heavy paramagnetic particle free to rotate in space at finite temperature.[21]
4) A classical point electric dipole oscillator uniformly accelerated through
classical electromagnetic zero-point radiation behaves as though it were at
rest in the Planck spectrum of thermal radiation including zero-point
radiation.[22] 5) A classical spinning point magnetic moment uniformly
accelerated through classical electromagnetic zero-point radiation also
behaves as through it were at rest in the Planck spectrum with zero-point radiation.[23]

\noindent\textbf{CLASSICAL ELECTRON THEORY WITH CLASSICAL ELECTROMAGNETIC
ZERO-POINT RADIATION}

All of these results involving classical thermal radiation and classical
zero-point radiation point beyond the blackbody problem to the larger
questions of atomic physics. They suggest that one consider classical electron
theory with classical electromagnetic zero-point radiation, a theory which is
often termed "stochastic electrodynamics."[24] Indeed, for point electric
dipole oscillators, such calculations have been presented repeatedly[25] with
results which show satisfying agreement with quantum oscillator systems.
Furthermore, the correspondence holds well for all point mechanical systems
which have no harmonics. \ Thus the inclusion of classical electromagnetic
zero-point radiation in classical physics can provide quantitative
explanations of van der Waals forces, diamagnetism, and the specific heats of
solids as well as semiquantitative understanding of the absence of atomic collapse.[18]

\ Despite these significant successes, many of those interested in classical
analyses of atomic phenomena have despaired of an eventual classical
understanding. \ The confounding analysis appears in attempts to understand
thermal equilibrium in the presence of a classical electromagnetic scatterer.
All of the work to date,[26] including the heroic calculations of Blanco,
Pesquera, and Santos,[27] suggests that the Rayleigh-Jeans spectrum is the
unique spectrum of random radiation which is invariant under scattering by a
classical electromagnetic system which has harmonics. Such a conclusion dooms
a classical point of view.

Such a gloomy outlook may not be justified. \ It is the author's opinion that
theoretical physics continues to suffer from a blindness to the mismatch
between mechanics and electromagnetism which was noted in the nineteenth
century. The ideas of special relativity and of relativistic mechanics arose
at the beginning of the twentieth century from a recognition of this mismatch.
Quantum mechanics began at the same time as an ad hoc system to accommodate
this mismatch while still treating all mechanical systems as legitimate. The
inclusion of classical zero-point radiation as a boundary condition on
Maxwell's equations is one of the ideas which is required in any attempt to
understand Nature in terms of classical theory. However, two other ideas are
also surely needed. The first is the restriction to classical systems which
share the scaling properties of electromagnetic theory. \ The second is an
understanding of fluctuations at a level beyond the quasi-Markov approximation.

\noindent\textbf{SCALING OF BLACKBODY RADIATION}

Classical electromagnetic theory governed by Maxwell's equations contains no
intrinsic length, time, or energy; the theory is scale invariant and indeed
conformal invariant.[28] However, the scales of length, time, and energy are
all coupled together.[29] If the scale of one of these is changed, then, the
scales of the others must change. Thus the scales of length and time are
connected through the speed of light in vacuum $c$. The scales of energy and
length are coupled through Coulomb's law for particles of unique charge $e$.

Nonrelativistic classical mechanics has no such connections between scales;
rather the scales of length, time, and energy can be chosen completely
independently. Relativistic classical mechanics moves toward the
electromagnetic situation since it introduces a coupling between the scales of
length and time through the special velocity $c$ which is the limiting speed
of a massive particle. However, within all of classical mechanics, energy
still scales completely separately from length and time.

The extremes of thermal radiation show the same contrast in energy coupling
and decoupling as noted here. The spectrum of zero-point radiation
$\mathcal{U}_{zp}(\omega)=(1/2)\hbar\omega$\ is invariant under a scale
transformation which couples length, time, and energy.[30] Such a
transformation preserves the values of both the velocity of light in vacuum
$c$ and the constant factor ($\hbar$) which gives the scale of the zero-point
spectrum. In contrast, the energy-equipartition Rayleigh-Jeans spectrum
$\mathcal{U}_{RJ}(T)=k_{B}T$\ completely decouples the thermal energy scale
from the length and time scales. Thus it is perhaps not surprising that
scattering calculations relying on classical mechanical systems, which do not
share the scaling properties of purely electromagnetic systems, should produce
only the scale-decoupled Rayleigh-Jeans spectrum.

\noindent\textbf{SCATTERING BY PURELY ELECTROMAGNETIC SYSTEMS}

Now back in 1989, when the scaling properties of blackbody radiation were
analyzed, it was suggested that only purely electromagnetic systems of charged
particles were suitable scatterers leading to equilibrium for classical
thermal radiation.[29] Purely electromagnetic systems allow conformal symmetry
where only the point masses must be transformed. Furthermore, the zero-point
radiation spectrum is the unique spectrum of random classical radiation which
is conformal invariant.[30] It is noteworthy that to date none of the
classical scattering calculations for radiation equilibrium has involved the
Coulomb potential or purely electromagnetic systems. Indeed, the previous
calculations place constraints on the allowed systems which specifically
exclude these systems from consideration.[31] The blackbody spectrum of
thermal radiation makes a smooth connection between the scale-coupled and the
scale-decoupled extremes, and one may suspect that only a purely
electromagnetic scattering system will show that the Planck spectrum with
zero-point radiation is the equilibrium spectrum of random classical radiation.[32]

Indeed, some unrelated classical electromagnetic work completed recently again
reminds us that purely electromagnetic systems can behave totally differently
from general mechanical systems. The interaction of a point charge and a
magnetic moment is completely different when the magnetic moment is modeled as
a charged particle in a Coulomb potential compared to when the magnetic moment
is modeled as a charged particle on a rigid ring.[33] It is only the purely
electromagnetic system which gives the foundation for understanding the
Aharonov-Bohm phase shift in terms of classical electromagnetic forces.[34]

\noindent\textbf{CONCLUDING SUMMARY OF THE ANALYSIS}

In this article, we explore the intimate connections between thermal radiation
and zero-point radiation within classical physics. \ First of all we review
the thermodynamics of a classical harmonic oscillator. Using the invariance of
the entropy under adiabatic changes of oscillator frequency, we find the
familiar result for the average oscillator energy $\mathcal{U}$ at (angular)
frequency $\omega$ and temperature $T$, $\mathcal{U}=\omega f(\omega/T)$ where
$f$ is an arbitrary function. This corresponds to Wien' displacement theorem.
However, all the derivations of this theorem in the textbooks involve moving
pistons in a cylindrical cavity rather than emphasizing that this is a result
holding for any system having a harmonic oscillator Hamiltonian. We then list
the associated forms of all the thermodynamic functions for a harmonic
oscillator system.

Next we discuss the high- and low-temperature limits which make the average
oscillator energy $\mathcal{U}$ independent of one of its two variables,
$\omega$ or $T$. Thus we recognize the equipartition result and the zero-point
result as the two extremes allowed by thermodynamics. Since zero-point
radiation is present in nature, we assume that the equilibrium thermodynamic
behavior of a harmonic oscillator moves monotonically between $\mathcal{U}%
=k_{B}T$ at low frequency and $\mathcal{U}=(1/2)\hbar\omega$ high frequency.
It is the problem of classical theory to determine the actual interpolation
formula. It is suggested that the true thermal spectrum corresponds to the
smoothest possible interpolation between the scale-decoupled equipartition
energy at low frequency and the scale-invariant zero-point energy at high
frequency. This suggestion leads to the Planck spectrum with zero-point energy.

At this point we turn from the thermodynamics of a single oscillator over to
the thermodynamics of radiation. For the sake of simplicity, we consider waves
in one space dimension. We derive the Stefan-Boltzmann law applicable in the
case of one space dimension, and we note that it gives us no information about
the radiation spectrum. Next we introduce a movable conducting partition
inside a box with conducting walls. The conducting boundary condition gives
rise to Casimir energies and Casimir forces associated with the changes in the
discrete spectrum of normal modes associated with the partition position.
Every different spectrum of random radiation leads to different Casimir forces
and energies. The zero-point spectrum is unique in that there is no entropy
change as the partition is moved within the conducting box. The equipartition
Rayleigh-Jeans spectrum is unique in that the energy of the box has no
dependence on the location of the conducting partition. Once again we suggest
that Nature would choose the blackbody radiation spectrum as the "smoothest
possible" interpolation between these extremes. We give a quantitative
criterion for the "smoothest possible interpolation" and report that of all
the monotonic energy spectra we have considered, the Planck spectrum is the
smoothest. We conjecture that analytic calculation would show that it is the
smoothest of all possible spectra. We emphasize that in the absence of
classical zero-point radiation, our criterion collapses and the traditional
"ultraviolet catastrophe" ensues. It is precisely the presence of zero-point
radiation which prevents the shift of thermal radiation to ever higher frequencies.

There are several other aspects of Casimir forces which are noted. We find
that although the Rayleigh-Jeans spectrum involves no energy changes with the
partition position at fixed temperature, it still involves Casimir forces on
the partition and hence changes in the Helmholtz free energy. The changes in
the Helmholtz free energy are tied to temperature-independent changes in the
entropy associated with the placement of the partition. Such
temperature-independent entropy changes are reminiscent of mixing entropy in
traditional classical statistical mechanics.

Although the calculations presented here involve waves in one spatial
dimension, the ideas can be carried over to electromagnetic waves in three
spatial dimensions.

Continuing our theme of the connections between blackbody radiation and
thermal radiation, we mention several derivations of Planck's spectrum in the
literature of classical physics, all of which depend upon classical
electromagnetic zero-point radiation.

Finally we discuss the failure of classical scattering calculations to give
anything but the Rayleigh-Jeans spectrum as an equilibrium spectrum. Working
from the scaling properties of thermal radiation and of electromagnetism, we
suggest that only purely electromagnetic systems have the appropriate scaling
properties to give the Planck spectrum with zero-point radiation as the
equilibrium spectrum. \ 

\noindent\textbf{CLOSING STATEMENT}

The blackbody radiation spectrum is a fundamental physical quantity which has
been of interest to physicists since the nineteenth century. The study of this
spectrum was crucial in the development of quantum mechanics. However,
discussion of blackbody radiation is disappearing from the text books.[35]
Indeed, today many physicists regard quantum theory as so firmly established
that it can be taught from axioms without regard to the physical phenomena
which caused the historical development of the theory. Nevertheless, there are
still some physicists who believe that the foundations of modern physics,
particularly in regard to classical electromagnetic zero-point radiation, have
never been adequately explored. At the same time, today's physicists are
sometimes being asked to swallow bizarre new "quantum" explanations simply
because physicists are unaware of the results of accurate classical
electromagnetic theory. The Aharonov-Bohm and Aharonov-Casher phase shifts, on
which the author has worked in recent years, seem prime examples; the
"force-free quantum topological effects" which are given as "explanations" for
these phenomena are accepted simply because of physicists' ignorance of
accurate classical electromagnetic predictions.[34]

It is suggested that twentieth century classical physics suffered from three
fundamental failures. i)The failure to include classical electromagnetic
zero-point radiation in the boundary conditions on Maxwell's equations. ii)The
failure to restrict analysis to purely electromagnetic systems when dealing
with fundamentally electromagnetic phenomena. iii)The failure to go beyond the
quasi-Markoffian approximation when dealing with systems in random classical
electromagnetic radiation. It is hoped that some time in the future these
failures will be rectified, and then we will indeed know the predictions of
classical physics in the atomic domain

\textbf{ACKNOWLEDGEMENT }

I wish to thank Professor Daniel C. Cole for his kind invitation to present a
talk on historical aspects of the Casimir model of the electron at the Seventh
International Conference on Squeezed States and Uncertainty Relations held at
Boston University during June 2001. The beginnings of this work arose out of
communications with Professor Cole.


\begin{thebibliography}{99}                                                                                               %


\bibitem {1}See, for example, H. Goldstein, \textit{Classical Mechanics 2nd
ed} (Addison-Wesley, Reading, MA, 1980), p. 462. We are using the notation
$\tilde{J}=J/(2\pi)$, $\tilde{w}=2\pi w$, where $J$ and $w$ are the
action-angle variables used in this text.

\bibitem {2}See, for example, ref. 1, pp. 531-540. \ The situation involving
adiabatic change in frequency is most famous in the case of a pendulum length
which is changed slowly. See, for example, Goldstein's exercise 10, p. 543.

\bibitem {3}See, for example, M. Planck, \textit{The Theory of Heat Radiation}
(Dover, NY 1959), pp. 72-83, or F. K. Richtmyer, E. H. Kennard, and T.
Lauritsen, \textit{Introduction to Modern Physics} (McGraw-Hill, New York
1955), pp. 113-118. \ 

\bibitem {4}C. Garrod, \textit{Statistical Mechanics and Thermodynamics}
(Oxford, New York 1995), p. 128.

\bibitem {5}M. J. Sparnaay, "Measurement of the attractive forces between flat
plates," Physica \textbf{24}, 751-764 (1958); S. K. Lamoreaux, "Demonstration
of the Casimir force in the 0.6 to 6$\mu$m range," Phys. Rev. Lett.
\textbf{78}, 5-8 (1997), \textbf{81}, 5475-5476 (1998); U. Mohideen,
"Precision measurement of the Casimir force from 0.1 to 0.9 $\mu$m," Phys.
Rev. Lett. \textbf{81}, 4549-4552 (1998); and H. B. Chan, V. A. Aksyuk, R. N.
Kleiman, D. J. Bishop, and F. Capasso, "Quantum mechanical actuation of
microelectomechanical systems by the Casimir force," Science \textbf{291},
1941-1944 (2001).

\bibitem {6}The choice $\hbar=1$ is familiar to particle physicists. \ The
measurement of temperature in energy units is familiar in thermodynamics where
our choice corresponds to the use of what is usually termed $\tau$ instead of
$T$. \ See, for example, C. Kittel, \textit{Elementary Statistical Physics}
(Wiley, New York 1958), p. 27.

\bibitem {7}In this article we have discussed the case of waves in one spatial
dimension. However, the same thermodynamic analysis applies immediately in
three dimensions. The behavior of Casimir forces within a three-dimensional
rectangular conducting box with a conducting partition can be shown
numerically to repeat the same sort of behavior as found in the
one-dimensional case. \ Indeed, related calculations were done decades ago in
the three-dimensional calculations of M. Fierz, "Zur Anziehung leitender
Ebenen im Vacuum, "Helvetica Physica Acta \textbf{33}, 855-858 (1960), and of
T. H. Boyer, "Some Aspects of Quantum Electromagnetic Zero-Point Energy and
Retarded Dispersion Forces," Harvard doctoral thesis 1968 (unpublished),
particularly Fig. 4.

\bibitem {8}This force expression is consistent with the generalized force
given below Eq.(4). \ 

\bibitem {9}See, for example, M. Planck, \textit{The Theory of Heat Radiation}
(Dover, NY 1959), pp. 61-63, or R. Becker and G. Leibfried, \textit{Theory of
Heat 2nd ed.} (Springer, NY 1967), pp. 16-17, or P.M. Morse, \textit{Thermal
Physics 2nd ed} (Benjamin/Cummings, Reading, MA 1969), pp. 78-79.

\bibitem {10}See, for example, E. A. Power, \textit{Introductory Quantum
Electrodynamics} (American Elsevier, NY 1964), pp. 18-22.

\bibitem {11}H. B. G. Casimir, Proc. Kon. Ned. Akad. Wetenschap. \textbf{51},
793 (1948) gives the force per unit area due to electromagnetic zero-point
radiation. \ The Rayleigh-Jeans spectrum gives a different force per unit
area, $F/A=-\zeta(3)k_{B}T/(4\pi x^{3})$. \ See, for example, T. H. Boyer,
"Temperature dependence of Van der Waals forces in classical electrodynamics
with classical electromagnetic zero-point radiation," Phys. Rev. A
\textbf{11}, 1650-1663 (1975).

\bibitem {12}See for example, G. H. Hardy, \textit{Divergent Series} (Oxford
University Press, London, 1956).

\bibitem {13}It is curious and perhaps significant that the Euler-Maclaurin
expansion which enters Casimir calculations involves the same Bernoulli
numbers as appear in the coefficients of the hyperbolic tangent function. See,
for example, R. P. Boas and C. Stutz, "Estimating sums with integrals," Am. J.
Phys. \textbf{39}, 745 (1971) and M. Abramowitz and J. Stegun, eds.,
\textit{Handbook of Mathematical Functions} (Dover, New York, 1965), pp. 804
and 806.

\bibitem {14}J. C. da Silva, A. Matos Neto, H. Q. Placido, M. Revzen, and A.
E. Santan, "Casimir effect for conducting and permeable plates at finite
temperature," Physica A \textbf{292}, 411-421 (2001).

\bibitem {15}H. A. Lorentz, \textit{The Theory of Electrons} (Dover, New York,
1952). \ This is a republication of the 2nd edition of 1915. \ Note 6, p. 240,
gives Lorentz's explicit assumption on the boundary condition.

\bibitem {16}The idea of classical zero-point radiation has occurred
repeatedly. \ The earliest extensive suggestion is given by W. Nernst,
"\"{U}ber einen Versuch, von quantentheoretischen Betrachtungen zur Annahme
stetiger Energie \"{a}nderungen zur\"{u}ckzukehren," Verhandlungen der
Deutschen Physikalischen Geselschaft, \textbf{18}, 83-116 (1916). Nernst was
struck by the invariance of the zero-point spectrum under adiabatic
compression. \ The Lorentz\ invariance of the spectrum was noted by T. W.
Marshall, "Statistical electrodynamics," Proc. Cambridge Phil. Soc.
\textbf{61}, 537-546 (1965), and later by others.

\bibitem {17}Far too many foolish referees see Planck's constant and conclude
that quantum mechanics is being used somehow. Indeed, there are no quanta
whatsoever in the classical analysis of the present paper. By the symbol
$\hbar$, we actually mean $[\pi^{2}k_{B}^{4}/(15c^{3}a)]^{1/3}$ where $c$ is
the speed of light, $k_{B}$ is Boltzmann's constant, and $a$ is Stefan's
energy-density constant, all constants of classical physics. Stefan's constant
$a$ was introduced into physics in 1879, long before any suggestions of quanta.

\bibitem {18}L. de la Pena and A. M. Cetto have provided a extensive review of
work on classical electromagnetic zero-point radiation in \textit{The Quantum
Dice: An Introduction to Stochastic Electrodynamics} (Kluwer, Boston 1996).
For a short review, see also T. H. Boyer, "Random electrodynamics: The theory
of classical electrodynamics with classical electromagnetic zero-point
radiation," Phys. Rev. D 11, 790-808 (1975). \ 

\bibitem {19}T. H. Boyer, "Derivation of the blackbody radiation spectrum
without quantum assumptions," Phys. Rev. \textbf{182}, 1374 (1969).

\bibitem {20}T. H. Boyer, "Classical statistical thermodynamics and
electromagnetic zero-point radiation," Phys. Rev. \textbf{186}, 1304-1318 (1969).

\bibitem {21}T. H. Boyer, "Diamagnetism of a free particle in classical
electron theory with classical electromagnetic zero-point radiation," Phys.
Rev. A \textbf{21}, 66-72 (1980); "Derivation of the Planck radiation spectrum
as an interpolation formula in classical electrodynamics with classical
electromagnetic zero-point radiation," Phys. Rev. D \textbf{27}, 2906-2911
(1983); \textbf{29}, 2418-2419 (1984).

\bibitem {22}T. H. Boyer, "Thermal effects of acceleration for a classical
dipole oscillator in classical electromagnetic zero-point radiation," Phys.
Rev. D \textbf{29}, 1089-1095 (1984); "Derivation of the blackbody radiation
spectrum from the equivalence principle in classical physics with classical
electromagnetic zero-point radiation," Phys. Rev. D \textbf{29}, 1096-1098 (1984).

\bibitem {23}T. H. Boyer, "Thermal effects of acceleration for a classical
spinning magnetic dipole in classical electromagnetic zero-point radiation,"
Phys. Rev. D \textbf{30}, 1228-1232 (1984).

\bibitem {24}Ideas of classical electromagnetic zero-point radiation appear
extensively in the work of T. W. Marshall, T. H. Boyer, D. C. Cole, and
others. See the review by L. de la Pena and A. M. Cetto listed in reference 18.

\bibitem {25}Among the earlier calculations is that of P. Braffort and C.
Tzara, "Energie de l'osillateur harmonique dans le vide," Compte Rendu Acad.
Sci. Paris \textbf{239}, 1779-1780 (1954). \ See Chapter 7 in the review by L.
de la Pena and A. M. Cetto listed in reference 18.

\bibitem {26}J. H. van Vleck and D. L. Huber, Rev. Mod. Phys \textbf{49}, 939
(1977). T. H. Boyer, "Equilibrium of random classical electromagnetic
radiation in the presence of a nonrelativistic nonlinear electric dipole
oscillator," Phys. Rev. D \textbf{13}, 2832-2845 (1976); "Statistical
equilibrium of nonrelativistic multiply periodic classical systems and random
classical electromagnetic radiation," Phys. Rev. A \textbf{18}, 1228-1237 (1978).

\bibitem {27}R. Blanco, L. Pesquera, and E. Santos, "Equilibrium between
radiation and matter for classical relativistic multiperiodic systems.
Derivation of Maxwell-Boltzmann distribution from Rayleigh-Jeans spectrum,"
Phys. Rev. D \textbf{27}, 1254-1287 (1983); "Equilibrium between radiation and
matter for classical relativistic multiperiodic systems. II. Study of
radiative equilibrium with Rayleigh-Jeans radiation," Phys. Rev. D
\textbf{29}, 2240-2254 (1984).

\bibitem {28}E. Cunningham, "The principle of relativity in electrodynamics
and an extension thereof," Proc. London Math. Soc. \textbf{8}, 77-98 (1910);
H. Bateman, "The transformation of the electrodynamical equations," Proc.
London Math. Soc. \textbf{8}, 223-264 (1910).

\bibitem {29}T. H. Boyer, "Scaling symmetry and thermodynamic equilibrium for
classical electromagnetic radiation," Found. Phys. \textbf{19}, 1371-1383 (1989).

\bibitem {30}T. H. Boyer, "Conformal symmetry of classical electromagnetic
zero-point radiation," Found. Phys. \textbf{19}, 349-365 (1989).

\bibitem {31}See references 26 and 27. It seems noteworthy that the
quasi-Markov limit required in the available scattering calculations leads to
singular behavior for electromagnetic systems.

\bibitem {32}Note that classical electron theory fits naturally into the
curved spacetime of general relativity whereas most mechanical systems do not.
This is related to the conformal symmetry of Maxwell's equations.

\bibitem {33}T. H. Boyer, "Classical Electromagnetic Interaction of a Point
Charge and a Magnetic Moment: Considerations Related to the Aharonov-Bohm
Phase Shift," Found. Phys. \textbf{32}, 1-39 (2002).

\bibitem {34}T. H. Boyer, "Semiclassical Explanation of the Matteucci-Pozzi
and Aharonov-Bohm Phase Shifts;" Found. Phys. \textbf{32}, 41-49 (2002); "Does
the Aharonov-Bohm Effect Exist?" Found. Phys. \textbf{30}, 893 (2000).

\bibitem {35}For example, blackbody radiation does not appear in J. J.
Sakurai's text, \textit{Modern Quantum Mechanics (Revised Edition)}
(Addison-Wesley, Reading, MA, 1994), nor in L. E. Ballentine's text,
\textit{Quantum Mechanics }(Prentice Hall, Englewood Cliffs, NJ, 1990).
\end{thebibliography}
\end{document}